\documentclass[a4paper,12pt]{article}
\usepackage{graphicx, color, pifont,psfrag}
\usepackage{amsmath,amssymb,a4wide,epsfig}
\def\bs{\begin{slide}}\def\es{\end{slide}}
\def\bc{\begin{center}}\def\ec{\end{center}}
\def\bmini#1{\begin{minipage}{#1}}\def\emini{\end{minipage}}
\def\be{\begin{equation}}\def\ee{\end{equation}}
\def\bea{\begin{eqnarray}}\def\eea{\end{eqnarray}}
\def\lsim{\raise0.3ex\hbox{$\;<$\kern-0.75em\raise-1.1ex\hbox{$\sim\;$}}}
\def\gsim{\raise0.3ex\hbox{$\;>$\kern-0.75em\raise-1.1ex\hbox{$\sim\;$}}}

\def\vtb{$V_{tb} $}\def\tp{t^{\prime}}

\newcommand{\abs}[1]{\ensuremath{\left|#1\right|}}
\newcommand{\Vtb}{\ensuremath{\abs{V_{tb}}}}
\newcommand{\Vtd}{\ensuremath{\abs{V_{td}}}}
\newcommand{\Vts}{\ensuremath{\abs{V_{ts}}}}

\begin{document}
\begin{titlepage}
\begin{flushright}{CP3-06-01}\end{flushright}
\vspace*{2cm}

\bc{\Large Is $V_{tb}\simeq 1$? }
\vspace*{2cm}

{\large
J. Alwall, R. Frederix, J.-M. G\'erard, A. Giammanco, M. Herquet,\\ 
\medskip S. Kalinin, E. Kou, V. Lema\^itre, F. Maltoni}
\ec

\bigskip

\bc
{Centre for Particle Physics and Phenomenology (CP3) \\
Universit\'{e} Catholique de Louvain\\
Chemin du Cyclotron 2 \\
B-1348 Louvain-la-Neuve, Belgium
}\\
\ec
\bigskip
\center{\today}
\bigskip

\begin{abstract}
The strongest constraint on $V_{tb}$ presently comes from the $3 \times
3$ unitarity of the CKM matrix, which fixes $V_{tb}$ to be very close to one. 
If the unitarity is relaxed, current information from top
production at Tevatron still leaves open the possibility that $V_{tb}$
is sizably smaller than one. In minimal extensions of the standard
model with extra heavy quarks, the unitarity constraints are much weaker
and the EW precision parameters entail the strongest bounds on
$V_{tb}$. We discuss the experimental perspectives of discovering and
identifying such new physics models at the Tevatron and the LHC, 
through a precise measurement of $V_{tb}$ from the single top 
cross sections and by the study of processes where the 
extra heavy quarks are produced.
\end{abstract}

\end{titlepage}
\newpage
\section{Introduction}

The value of the CKM matrix element $V_{tb}$, related to the top-bottom
charged current, is often considered to be known to a very satisfactory 
precision ($0.9990<\Vtb<0.9992$ at 90\%
C.L.~\cite{Eidelman:2004wy}). However, this range is determined  using
a full set of tree-level processes and relies on the unitarity of the
$3\times3$ CKM matrix.  The unitary assumption is mainly supported by
three experimental facts:
\begin{enumerate}
 \item The measurement of $V_{ub}$ and $V_{cb}$ in $B$ mesons decays. 
We now know that the hierarchy of the elements belonging to 
the first two rows of the CKM
matrix is in excellent agreement with the unitary condition. This is
particularly evident within the Wolfenstein's parametrization in terms of
$\lambda\equiv \sin\theta_c\simeq 0.22$ where $\theta_c$ is the Cabibbo angle.
\item 
The recent D\O\ and CDF results  on $\Delta M_{B_s}$~\cite{Abazov:2006dm, moriond}:
\bea
17 \ \mbox{ps}^{-1} < \Delta M_{B_s} < 21\  \mbox{ps}^{-1} \quad {\mbox{(90\% C.L. interval)}} \quad &&\mbox{D\O\ collaboration }\quad \\
17.33^{+0.42}_{-0.21}{\rm (stat.)}\pm 0.07{\rm (syst.)}\  \mbox{ps}^{-1}\quad \ \quad && \mbox{CDF collaboration }. 
\eea
The rather precise CDF measurement allows us to extract the ratio $|V_{td}/V_{ts}|$ 
\be
0.20 < |V_{td}/V_{ts}| < 0.22\,, \label{eq:vtdvts}
\ee
by using  $\Delta M_{B_d}/\Delta M_{B_s}$ (see, {\it  e.g.}, Ref.~\cite{Eidelman:2004wy})  and taking into account the theoretical uncertainty  
associated with the hadronic matrix elements~\cite{okamoto}. 
This ratio fits well with the unitary hypothesis which predicts it to be of order $\lambda$.
One should emphasize however, that   these processes come from loop diagrams, and 
could be polluted by new physics contributions. 
\item The Tevatron measurements of $R$ based on the relative
number of $t\bar t$-like events with zero, one and two
tagged $b$-jets. The resulting values for $R$ are 
$1.12^{+0.27}_{-0.23}~\rm(stat.+syst.)$~\cite{Acosta:2005hr} and $1.03^{+0.19}_{-0.17}~\rm(stat.+syst.)$
\cite{unknown:2006bh} 
for CDF and D\O\ respectively, both giving $R > 0.61$ at 95\%
confidence level. Recalling  the definition
\begin{equation}
R\equiv\frac{\Vtb^2}{\Vtd^2+\Vts^2+\Vtb^2}, \label{eq:Rdef}
\end{equation}
it is clear that $R\simeq 1$ implies a strong
hierarchy between $V_{tb}$ and the other two matrix elements, 
as expected in the unitary case. As we will argue later on, 
the upper limits of the single top production cross sections from Tevatron might 
already  provide (rather loose) additional constraints on their absolute magnitude, 
$|V_{ts}|\lsim 0.62$ and  $|V_{td}|\lsim 0.46$.
\end{enumerate}

On the other hand, contrary to what has sometimes been argued, none of
these experimental facts are \textit{directly} constraining $V_{tb}$.
In fact, even its ``direct'' determination from $R$, giving
$\Vtb > 0.78$ at 95\% C.L., comes simply from taking the square root of $R$, 
assuming the unitarity of the CKM matrix.  Since no single top 
cross section measurement yet exists, the $V_{tb} \neq 1$ alternative
should be considered as still acceptable. This possibility
appears, for example, if one introduces new heavy up- and/or
down-type quarks. Though such new fermions are not favoured by current
precision constraints, they are not yet excluded, and their
existence is in fact predicted by many extensions of the Standard Model
(SM)~\cite{models}.  We should thus keep in mind that the familiar
$3\times3$ CKM matrix might well be a submatrix of a $3\times4$,
$4\times3$, $4\times4$ or even larger matrix.

In the following section, we present two minimal extensions of the SM
that allow a value for $V_{tb}$ considerably different from one. Although these
models are theoretically self-consistent, they should be primarily
regarded as motivations for further experimental scrutiny of $V_{tb}$.
In the first case, the introduction of a new vector-like top singlet
leads to a global rescaling of $V_{td}$, $V_{ts}$ and $V_{tb}$ leaving $R$
unchanged.  In the second case, a complete
new fourth generation is added and the $R$ measurement is used as a
direct constraint. In Section 3, we discuss the expected
precision on the extraction of $V_{tb}$ at the LHC from the measurement
of the single top production cross sections.  Finally, we review some
aspects of direct $t'$ search at the LHC and in particular the
possibility of distinguishing a vector-like $SU(2)_L$ singlet top
from that of a fourth generation.

\section{Models allowing sizable deviations from  $V_{tb}\simeq 1$}\label{sec:2}

\subsection{The case for a vector-like $\tp$ quark}\label{sec:2.1}
As discussed in the introduction, a ratio $R$ close to one does not necessarily require $V_{tb}$ to be close to one. Indeed, as can be seen from Eq.~(\ref{eq:Rdef}), this ratio is invariant under a  simple rescaling of all $V_{ti}^{(0)}$ entries: 
\be
V_{ti}=V_{ti}^{(0)} \cos\theta. 
\ee
The minimal way to implement such a rescaling within the so successful renormalizable $SU(2)_L\times U(1)$ electroweak theory is to introduce one $Q=+2/3$ vector-like quark. If this hypothetical iso-singlet quark also has  a mass around  the electroweak scale, it naturally mixes with its nearest neighbour, {\it  i.e.}, the standard heavy top, to enlarge the unitary CKM matrix $V_{3\times 3}^{(0)}$: 
\be
{\bf V}_{4\times 3}=
\left(\begin{array}{cc}
{\bf 1}_{2\times 2} & 0 \\
0& {\bf U}_{2\times 2}
\end{array}\right)
\left(\begin{array}{c}
{\bf V}_{3\times 3}^{(0)} \\
0
\end{array}\right); 
\quad {\bf V}{\bf V}^{\dagger}\neq {\bf 1}_{4\times 4}\,, \label{eq:6}
\ee
where ${\bf V}$ enters in the flavor changing charged current
\be
{\mathcal{L}}_{W^{\pm}}(\theta )=-\frac{g}{\sqrt{2}}[\bar{u}_L {\bf V} \gamma^{\mu}d_LW_{\mu}^+ +h.c. ]\,.
\label{eq:L1}\ee
Note that such an enlargement does not spoil the unitarity of the
first two rows of the CKM matrix.  If we neglect possible CP-violating
phases beyond CKM, the left-handed unitary transformation leading to
the physical $t$ and $\tp$ quarks is a simple rotation in the $3-4$
flavour plane
\be
{\bf U}=R_{34}(\theta)=\left(\begin{array}{cc}
\cos\theta  & -\sin\theta  \\
\sin\theta & \cos\theta 
\end{array}\right)
\ee
such that 
\bea
V_{ti}&=& V_{ti}^{(0)} \cos\theta, \\
V_{\tp i}&=& V_{ti}^{(0)} \sin\theta, 
\eea
with $V_{tb}^{(0)} \simeq 1$. 
We are therefore left with only two new parameters beyond the SM, namely the $t-\tp$ mixing angle $\theta$ and the $\tp$ mass $m_{\tp}$. 
These arise from the following $SU(2)_L\times U(1)$ invariant Yukawa interactions: 
\be
{\mathcal{L}}_y(\tp) =\lambda(\overline{t^0, b^0})_L \Phi t_R^0+\lambda^{\prime} (\overline{t^0, b^0})_L \Phi t^{\prime 0}_{R}+h.c.
\ee
and Dirac mass terms
\be
{\mathcal{L}}_D(\tp)=M\bar{t}_L^{\prime 0}t_R^{\prime 0}+M^{\prime}\bar{t}_L^{\prime 0}t_R^{ 0}+h.c.
\ee
Assuming the $\tp$ mass  to be dominated by the new scale $M$ and not by the vacuum expectation value $v$ of the SM Higgs doublet $\Phi$,  $\lambda^{(\prime)}v<M^{(\prime)}$, the mixing angle $\theta$ is naturally smaller than $\pi/4$ and a  theoretical bound on \vtb\  is obtained as:
\be
|V_{tb}|\simeq |\cos\theta | > 1/\sqrt{2}\simeq 0.71. \label{eq:theory}
\ee  
This model allows \vtb \ to be smaller than one but also implies tree-level flavour changing neutral currents  (FCNC)
\bea
{\mathcal{L}}_{Z^0}(\theta)&=&-\frac{g}{2\cos \theta_{\mbox{w}}} \bar{u}_L {\bf V} {\bf V}^{\dagger} \gamma^{\mu}u_L Z_{\mu}^0 \label{eq:Z-fcnc}\\
{\mathcal{L}}_{H^0}(\theta, m_{\tp})&=&\frac{g}{2M_W}[ \bar{u}_L {\bf V} {\bf V}^{\dagger} {\bf M}^u u_R +h.c.]H^0
\label{eq:L2}
\eea
with 
\be
{\bf V}{\bf V}^{\dagger}=\left(\begin{array}{ccc}
{\bf 1}_{2\times 2} &0 &0 \\
0&\cos^2\theta  &\sin\theta \cos\theta \\
0& \sin\theta \cos\theta& \sin^2\theta 
\end{array}\right); \quad {\bf M}^u={\rm diag}(m_u, m_c, m_t, m_{\tp}). \label{eq:Nmatrix}
\ee
Notice that the $Z$ coupling to $t\bar{t}$ is reduced by a factor of $\cos^2\theta$. 
The non-observation of the FCNC  processes potentially restricts  the off-diagonal elements of ${\bf V}{\bf V}^{\dagger}$ and consequently constrains the  $t-\tp$ mixing angle $\theta$. 
In fact, current limits on FCNC involving top quark only constrain the $Ztq$ couplings  ($q=u, c$)~\cite{Eidelman:2004wy}. 

We comment in passing on the similar model but with a down-type vector-like quark, $b^{\prime}$. 
In this case, the $3\times 4$ matrix can be written in terms of a single mixing angle $\theta_d$ by the transposed of the $4\times 3$ matrix in Eq.~(\ref{eq:6}) and $V_{tb}$ is now scaled as $V_{tb}=V_{tb}^{(0)}\cos\theta_d$.   
However, contrary to the $\tp$ case to which we shall come back in Section \ref{sec:2.1.2}, this angle is now very strongly constrained by the measurement of $R_b\equiv \Gamma(Z\to b\bar{b})/\Gamma(Z\to \mbox{hadrons})$ since the $Z$ coupling to $b\bar{b}$ is reduced by a factor of $\cos^2\theta_d$ at the tree level. One can write $R_b$ in terms of its SM prediction $R_b^{\mbox{\tiny SM}}$ as
\be
R_b\simeq R_b^{\mbox{\tiny SM}}[1-(1-R_b^{\mbox{\tiny SM}})\sin^2\theta_d]. 
\ee
The precisely known experimental and theoretical values constrain $\sin\theta_d$ to be smaller than 0.06, 
which leads to a maximum  reduction of $V_{tb}$ compared to $V_{tb}^{(0)}$ of only 0.2\%.

\subsubsection{Current constraints on $\tp$ mass}\label{sec:2.1.1}

Recently, a new result  with the 760 pb$^{-1}$ data of the CDF Run II was announced \cite{CDFRII}, which excludes a $\tp$ mass below 258 GeV at 95\% C.L. This limit is obtained by assuming the branching ratio of $\tp\to W^+q $ to be equal to unity. Thus, if $\tp$ had other decay channels, namely flavour changing neutral modes in our model, this bound would be less strict. 

\begin{figure}[t]
  \hfill
  \begin{minipage}[t]{.49\textwidth}
    \begin{center}  
      \epsfig{file=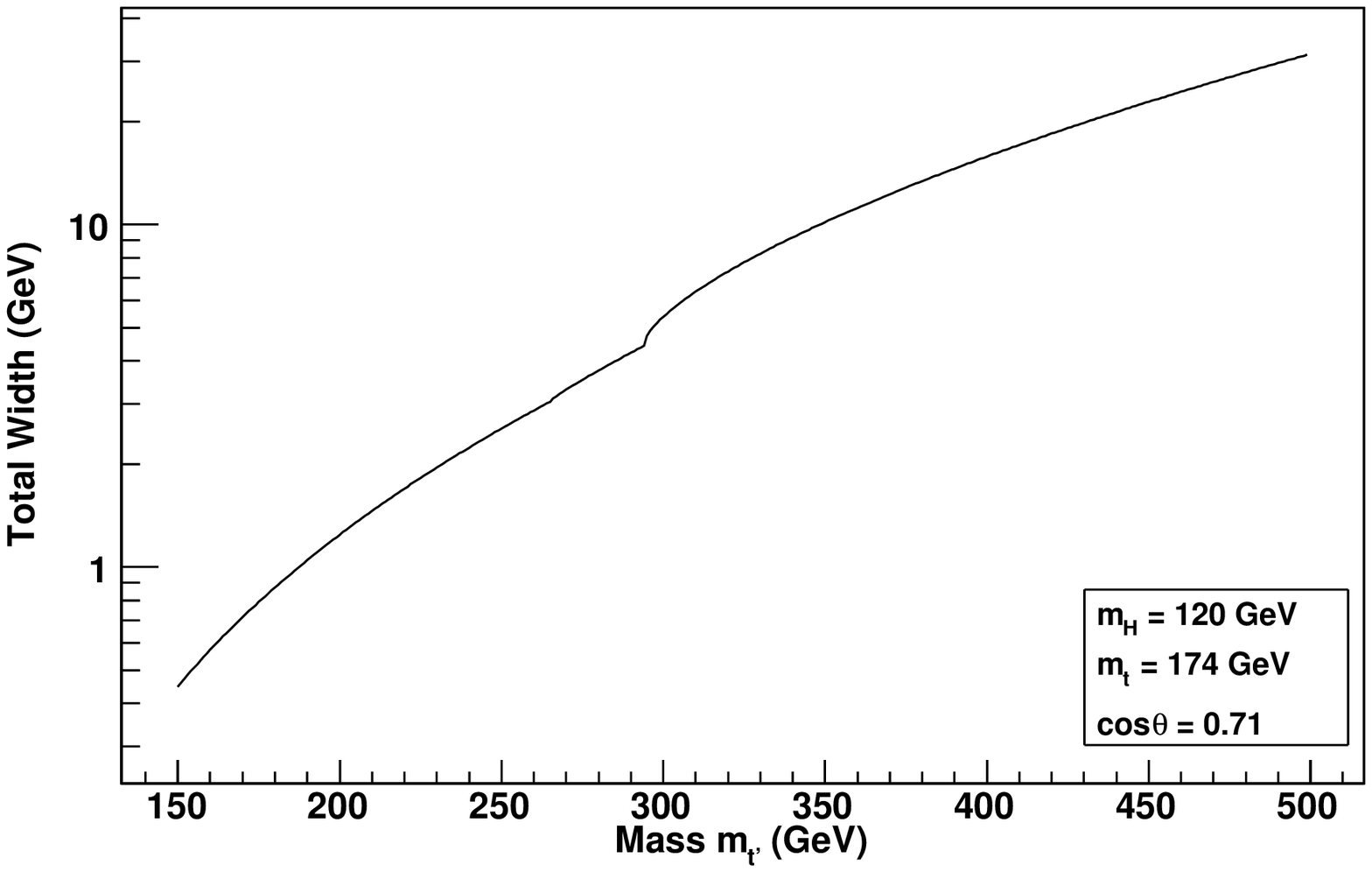, width=1.1\textwidth}\\
      (a)
    \end{center}
  \end{minipage}
  \hfill
  \begin{minipage}[t]{.49\textwidth}
    \begin{center}  
      \epsfig{file=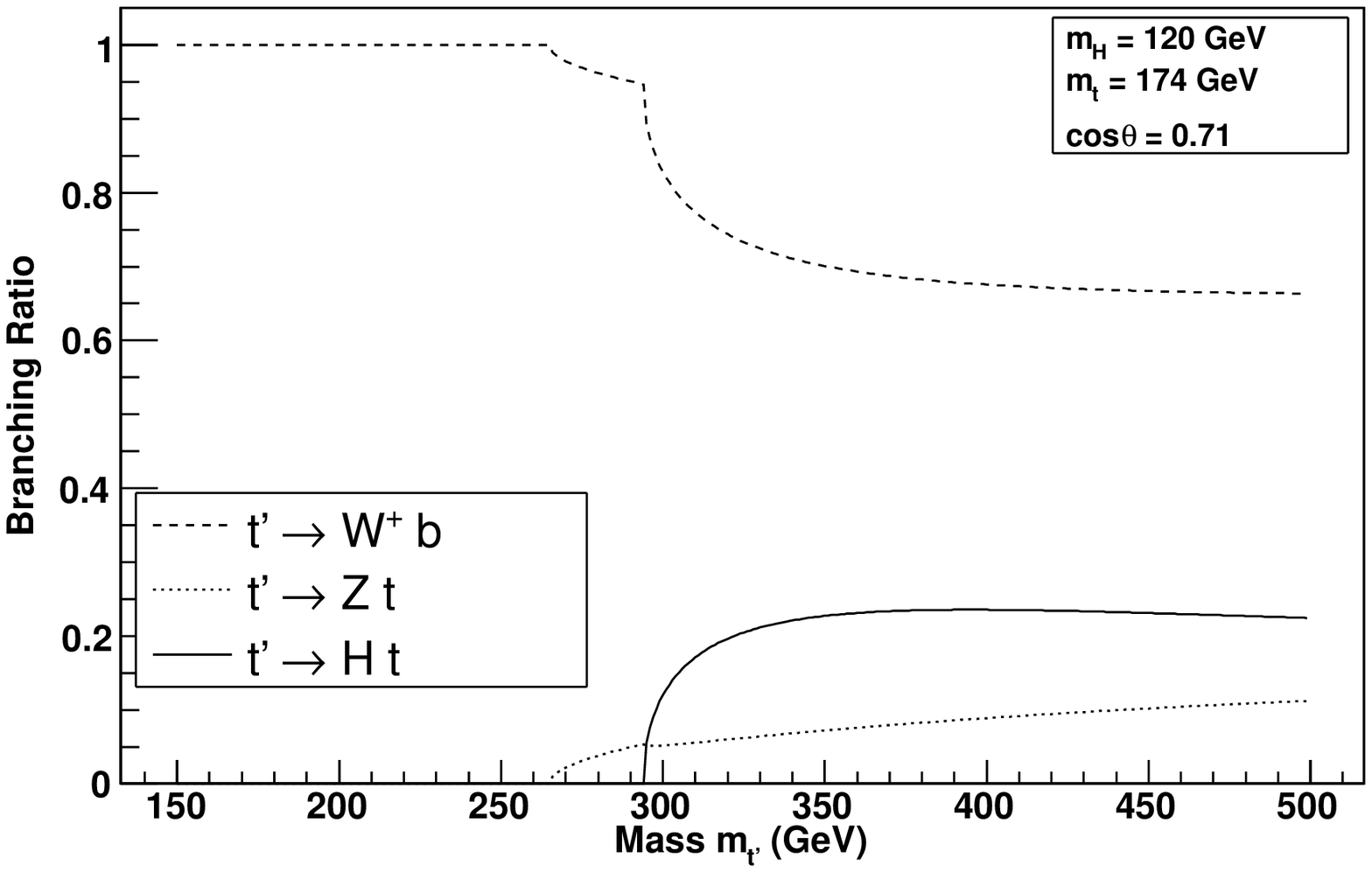, width=1.1\textwidth}\\
      (b)
    \end{center}
  \end{minipage}
  \hfill
  \caption{The total width (a) and the branching ratios (b) for the decay of the $t'$ as a function of the $t'$ mass. }
  \label{extra_top_BR}
\end{figure}
At leading order, $\tp$ has three decay modes,  $\tp\to W^+b$, $\tp\to Zt/Ht$ (see Eqs. (\ref{eq:L1}), (\ref{eq:Z-fcnc}) and  (\ref{eq:L2})). The on-shell decay widths are given by~\cite{Aguilar-Saavedra:2005pv}
\begin{align}
\Gamma(t'\to W^+b)&=\frac{\alpha}{16 s_W^2}\frac{m_{t'}^3}{m_W^2}|V_{bt'}|^2 (1+x_W-2x_W^2-2x_b+x_Wx_b+x_b^2)\sqrt{\lambda(1,x_W,x_b)},\nonumber\\
\Gamma(t'\to Z t)&=\frac{\alpha}{32s_W^2}\frac{m_{t'}^3}{m_W^2}|({\bf VV^{\dagger}})_{tt'}|^2(1+x_Z-2x_Z^2-2x_t+x_Zx_t+x_t^2)\sqrt{\lambda(1,x_Z,x_t)},\label{r2}\\
\Gamma(t'\to H t)&=\frac{\alpha}{32s_W^2}\frac{m_{t'}^3}{m_W^2}|({\bf VV^{\dagger}})_{tt'}|^2(1+6x_t-x_H+x_t^2-x_Hx_t)\sqrt{\lambda(1,x_H,x_t)},\nonumber
\end{align}
where
\begin{equation}\label{r4}
\lambda(1,x,y)=1+x^2+y^2-2x-2y-2xy, \quad x_i=\frac{m_i^2}{m_{t'}^2}. 
\end{equation}
The total decay width is given in Fig.~\ref{extra_top_BR}a while the branching ratios for the different
modes are given in Fig.~\ref{extra_top_BR}b as a
function of the mass of the $t'$. Here we have set  $\cos\theta = 0.71$
and the mass of the Higgs boson to $m_H=120$ GeV.

For $\tp$ masses below the $Z$-boson plus top quark threshold ($\sim
265$ GeV), the only on-shell decay is $\tp\to W^+b$.  For $\tp$
masses between $\sim 265$ GeV and $\sim 295$ GeV, there is also a small
contribution from the second mode in  Eq.~(\ref{r2}). For $\tp$ masses larger
than $\sim 295$ GeV, {\it i.e.}, the top and Higgs threshold, none of 
the three decay modes can be neglected.

For larger $\cos\theta$ the branching ratio $\textrm{Br}(t'\to W^+ b)$
will be reduced.  For example, for $\cos\theta= 0.9$ and a $\tp$ mass
larger than $\sim 375$ GeV more than 45\% of the decays will be
$\tp\to Zt / Ht$.  A larger Higgs boson mass will lower the branching
ratio $\textrm{Br}(t'\to H t)$.  Nevertheless, the current CDF bound
is not affected by those extra contributions. Thus, in the following
we use:
\be
\frac{m_{\tp}}{m_t}\geq 1.5 \quad \mbox{(95\% C.L.)\,.} \label{eq:tpmass}
\ee

\subsubsection{Current constraints on $t-\tp$ mixing}\label{sec:2.1.2}

\begin{figure}[t]\begin{center}
\psfrag{x}[c][c][0.9]{$\cos\theta$}\psfrag{y}[c][c][1]{\rotatebox{270}{$\frac{m_{\tp}}{m_t}$}}
\psfrag{one}[c][c][0.8]{1\ $\sigma$}\psfrag{ninty}[c][c][0.8]{90\% C.L.}\psfrag{nintyfive}[c][c][0.8]{95\% C.L.}
\psfrag{tpmass}[c][c][1]{\rotatebox{270}{$\leftarrow$} }
\psfrag{1.2}[c][c][0.7]{1.2}\psfrag{1.4}[c][c][0.7]{1.4}\psfrag{1.6}[c][c][0.7]{1.6}\psfrag{1.8}[c][c][0.7]{1.8}\psfrag{2}[c][c][0.7]{2.0\ \ }
\psfrag{-0.75}[c][c][0.7]{-0.75}\psfrag{-0.5}[c][c][0.7]{-0.50}\psfrag{-0.25}[c][c][0.7]{-0.25}
\psfrag{0}[c][c][0.7]{0}\psfrag{1}[c][c][0.9]{\ }
\psfrag{0.75}[c][c][0.7]{0.75}\psfrag{0.5}[c][c][0.7]{0.50}\psfrag{0.25}[c][c][0.7]{0.25}
\psfrag{tv1}[c][c][0.8]{{$\theta_v=0.2$}}\psfrag{tv2}[c][c][0.8]{{$\theta_v=0.1$}}
\psfrag{x}[c][c][0.9]{$|\cos\theta|$}
\psfrag{y}[c][c][1]{\rotatebox{270}{$\frac{m_{\tp}}{m_t}$}}
\psfrag{one}[c][c][0.8]{1\ $\sigma$}
\psfrag{ninty}[c][c][0.8]{90\% C.L.}
\psfrag{nintyfive}[c][c][0.8]{95\% C.L.}
\psfrag{tpmass}[c][c][1]{\rotatebox{270}{$\leftarrow$} }
\psfrag{1.2}[c][c][0.7]{1.2}
\psfrag{1.4}[c][c][0.7]{1.4}
\psfrag{1.6}[c][c][0.7]{1.6}
\psfrag{1.8}[c][c][0.7]{1.8}
\psfrag{2}[c][c][0.7]{2.0\ \ }
\psfrag{0.8}[c][c][0.7]{0.8}
\psfrag{0.6}[c][c][0.7]{0.6}
\psfrag{0.4}[c][c][0.7]{0.4}
\psfrag{0.2}[c][c][0.7]{0.2}
\psfrag{1}[c][c][0.7]{1.0}
\psfrag{a}[c][c][1]{\rotatebox{270}{$\leftarrow$} }
\psfrag{tv1}[c][c][0.8]{{$\theta_v=0.2$}}
\psfrag{tv2}[c][c][0.8]{{$\theta_v=0.1$}}
\begin{minipage}{8cm}
\includegraphics[width=7.5cm]{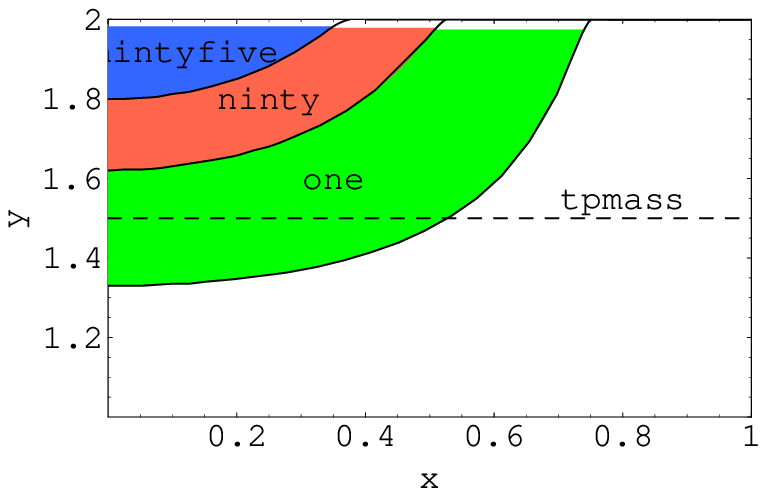}
\begin{center} (a) \end{center}\end{minipage}
\begin{minipage}{8cm}
\includegraphics[width=7.5cm]{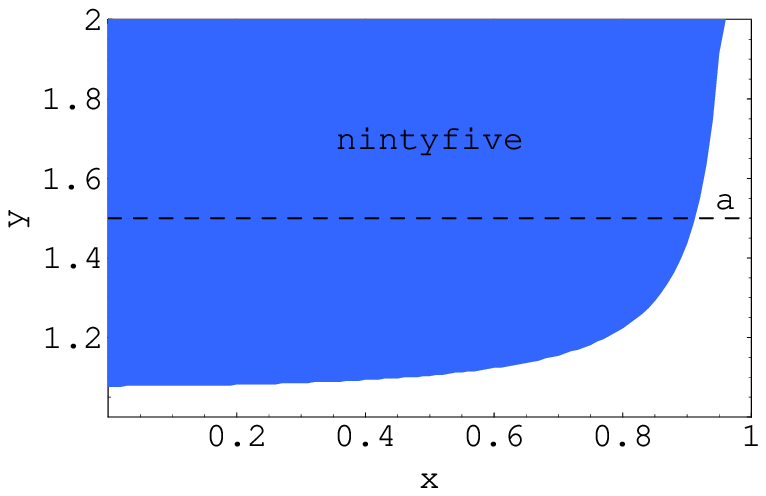}
\begin{center} (b) \end{center}\end{minipage}
\caption{Excluded range for the mass and mixing of a vector-like quark $\tp$
from $B\to X_s \gamma $  at 
95\%, 90\%, 68.3\%  C.L.  (a) and  $R_b$  at 95\% C.L. (b).  The horizontal dashed line 
indicates the experimental bound on $m_{\tp}$ at 95\% C.L., Eq.~(\ref{eq:tpmass}). }
\label{fig:bsgamma}\end{center}
\end{figure}

We now turn to the experimental constraints for $\theta$ and $m_{\tp}$. The strongest flavour physics constraint  comes from the branching ratio of $B\to X_s\gamma$. 
The correction to the amplitude of $B\to X_s \gamma$ scales like~\cite{buras-lindner}
\be
\left[\left(\frac{m_{\tp}}{m_t}\right)^{0.60}-1\right]\sin^2\theta\,,
\ee
if $m_{\tp}<300$ GeV. 
Computing the branching ratio at  NLO accuracy as in  Refs.~\cite{KN, CMM}, the allowed range for $\cos\theta$ from the precise measurement 
\be
{\rm Br}(B\to X_s\gamma )=(3.55\pm 0.45) \times 10^{-4}
\ee
leads to the constraints shown in Fig.~\ref{fig:bsgamma}a. Together with the constraint for $m_{\tp} $ 
in Eq.~({\ref{eq:tpmass}}), it translates into a lower bound for $|V_{tb}|$ with 
\begin{equation}
|\cos\theta|_{B\to X_s \gamma} > 0.53\,,
\end{equation} 
where only one $\sigma$ of experimental uncertainty  in ${\rm Br}(B\to X_s \gamma)$ is included. 
Notice that this bound is still weaker than the theoretical 
one coming from  Eq.~(\ref{eq:theory}). As can be seen in the figure, at a higher confidence level, we do not obtain any constraint on \vtb\  from $B\to X_s\gamma$.  

As a next step, we consider the  constraints coming from the electroweak precision measurements.  
The complete contribution of the $\tp$ particle to the $T$ parameter is positive and given by~\cite{LS} 
\be
T=\frac{3}{16 \pi \sin^{2}\theta_w\cos^2\theta_w}[\sin^2\theta F(y_{\tp}, y_b)-\sin^2\theta \cos^2\theta F(y_{t}, y_{\tp})-\sin^2\theta F(y_{t}, y_b)]\,,
\ee
where $y_i=m_i^2/m_Z^2$ and 
\be
F(y_1,y_2)=y_1+y_2-\frac{2y_1y_2}{y_1-y_2}\ln \frac{y_1}{y_2}; \quad F(y, y)=0.  
\ee
The experimental bound on $m_{\tp}$ in Eq.~(\ref{eq:tpmass}) implies
\be
T> 1.1 \sin^2\theta \quad \mbox{for} \ m_{\tp}>258\ \mbox{GeV}. \label{eq:TVLQ}
\ee
We find that the $S$ and $U$ parameters can be relatively small, $U>0.12 \sin^2\theta$ and  $S>-0.024 \sin^2\theta$, compared to $T$ in this model. 
A direct comparison with  the most recent experimental result from LEP \& SLD in~\cite{EWW},  $T=0.13\pm 0.10$, where Higgs mass is fixed to $m_H=150$ GeV, implies  $|\cos \theta| > 0.89$ if $T=0.23$. 
However, we would like to emphasize that the $T$ parameter is known to increase as the Higgs mass increases. Therefore, this constraint can be relaxed by including the uncertainties from the Higgs mass. 

\begin{center}
\begin{figure}[t]
  \hspace*{1.5cm}
  \begin{minipage}[t]{.40\textwidth}
      \epsfig{file=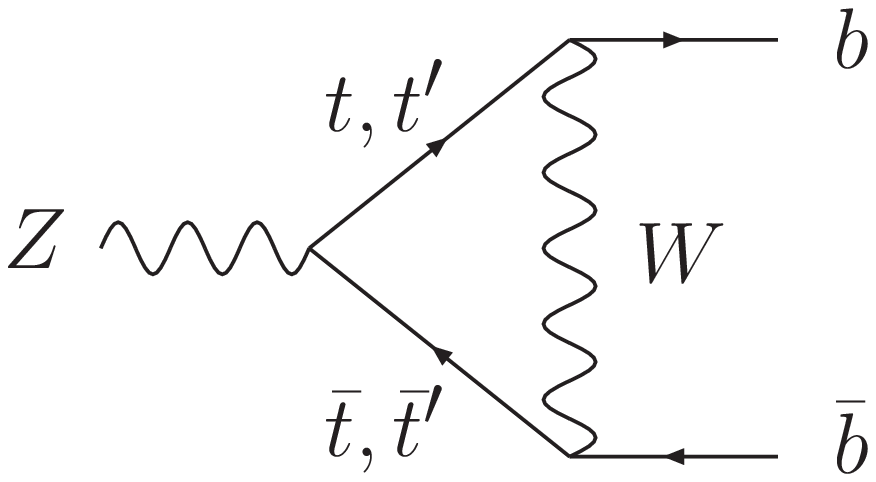, width=5cm}
  \end{minipage}
  \hspace*{0.2cm}
  \begin{minipage}[t]{.40\textwidth}
      \epsfig{file=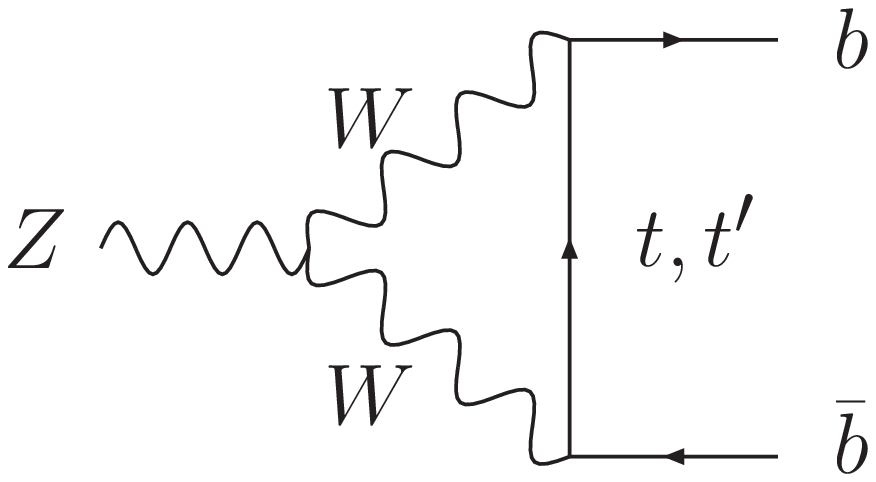, width=5cm}
  \end{minipage}
  \caption{Modification of the $Z\to b\bar b$ rate from one-loop diagram including $t$ and $\tp$. In the case of a vector-like $t'$, also the flavor changing neutral vertex $Ztt'$ contributes.}
  \label{fig:fynrb}
\end{figure}
\end{center}

On the other hand, the $R_b$ ratio, $\Gamma(Z\to b\bar{b})/\Gamma(Z\to \mbox{hadrons})$ turns out to give much stronger and more solid constraints. The $t$ and $\tp$ loop corrections to $\Gamma (Z\to b\bar{b})$ modify this ratio as (see Fig.~\ref{fig:fynrb})~\cite{LondonD}
\be
R_b\lsim (1-0.015 \sin^2\theta)R_b^{\rm SM}\,,\label{eq:rbvt}
\ee
if $m_{\tp}\gsim 258$ GeV is used. The current experimental result 
\be
R_b^{\rm exp} =0.21638\pm 0.00066
\ee
is consistent with the SM fitted value
\be
R_b^{\rm SM}=0.21564 \pm 0.00014
\ee
within $1.1\sigma$. Using 95\% C.L. value for the experimental data, we end up with a rather strong and solid constraint (see Fig.~\ref{fig:bsgamma}b), 
\be
|\cos\theta|_{R_b} \gsim 0.91.
\ee

\subsection{The case for a fourth generation}\label{sec:2.2}
Another possible extension of the CKM structure of SM
is the addition of a fourth generation. In this case, the presence of
$b^{\prime}$ implies a unitary $V_{4\times 4}$ mixing matrix such that
tree-level FCNCs in hadronic $Z^0$ decays are now forbidden (see
Eq.~(\ref{eq:Z-fcnc})).  Next, we shall discuss the $(V_{td}, V_{ts},
V_{tb})$ bounds for this model.

Neglecting again the CP-violating phases beyond CKM, the $4\times
4$ unitary matrix contains three extra mixings which we parametrize,
following Ref.~\cite{BC}, as 
\be {\bf V}_{4\times 4} = R_{34}(\theta_u)R_{24}(\theta_v)R_{14}(\theta_w)
\left(\begin{array}{cc}
    {\bf V_{3\times 3}^{(0)}}&{\bf 0}_{3\times 1}\\ {\bf 0}_{1\times 3}&1\\
\end{array}\right)\,,
\ee
where $R_{ij}(\theta)$ is the  rotation in the $i-j$ flavour plane. 
It is important to notice that for the $3\times 3$ unitarity matrix part,  ${\bf V_{3\times 3}^{(0)}}$,  the usual Wolfenstein's expansion is applicable irrespective to the size of $\theta_{u, v, w}$ in this particular parametrization. 
We then obtain (for $i=d, s, b$) 
\bea
V_{ui}&=& \cos\theta_wV_{ui}^{(0)} \label{eq:Vui}\\
V_{ci}&=& \cos\theta_vV_{ci}^{(0)} -\sin\theta_v \sin\theta_w V_{ui}^{(0)}\label{eq:Vci}\\
V_{ti}&=& \cos\theta_uV_{ti}^{(0)} -\sin\theta_u \sin\theta_v V_{ci}^{(0)}-\sin\theta_u \cos\theta_v \sin\theta_wV_{ui}^{(0)} \label{eq:Vtifour}\\
V_{\tp i}&=& \sin\theta_uV_{ti}^{(0)} +\cos\theta_u \sin\theta_v V_{ci}^{(0)}+\cos\theta_u \cos\theta_v \sin\theta_wV_{ui}^{(0)}. 
\eea
Using the fact that $(V_{ud}^{(0)}, V_{us}^{(0)}, V_{cd}^{(0)}, V_{cs}^{(0)})$ are written in terms of the  single parameter $\lambda$ up to ${\mathcal{O}}(\lambda^3)$, the $4\times 4$ unitarity condition  immediately constrains the two extra mixing angles appearing in Eqs.~(\ref{eq:Vui}) and (\ref{eq:Vci}).  The experimental values given in Ref.~\cite{Eidelman:2004wy} indeed imply 
\be
|\theta_w|\le {\mathcal{O}}(\lambda^2),  \quad |\theta_v|\leq {\mathcal{O}}(\lambda)\,. \label{eq:ffunit}
\ee

\subsubsection{Current constraints on $V_{ti}$}
Similarly to the vector-like model, the mixing angle $\theta_u$ is not
constrained from the unitarity condition since the third row is not
known. Given the hierarchy of Eq.~(\ref{eq:ffunit}), let us neglect $\theta_w$.
However, even a small value of $\theta_v$ could entail a large deviation of $V_{ti}$ from its SM value.  
By choosing maximal $t-\tp$  mixing, {\it i.e.}, $\theta_u=\pi/4$, 
Eq.~(\ref{eq:Vtifour}) reduces to; 
\be\begin{array}{lll}
  \sqrt{2}V_{td}=\underbrace{V_{td}^{(0)}}_{{\mathcal{O}}(\lambda^3)}
  + \sin\theta_v \underbrace{V_{cd}^{(0)}}_{{\mathcal{O}}(\lambda)},&
  \sqrt{2}V_{ts}= \underbrace{V_{ts}^{(0)}}_{{\mathcal{O}}(\lambda^2)}
  + \sin\theta_v \underbrace{V_{cs}^{(0)}}_{{\mathcal{O}}(1)},&
  \sqrt{2}V_{tb}= \underbrace{V_{tb}^{(0)}}_{{\mathcal{O}}(1)} +
  \sin\theta_v \underbrace{V_{cb}^{(0)}}_{{\mathcal{O}}(\lambda^2)}\,.
\end{array}\label{eq:fgmax}
\ee
We notice that $(|V_{td}|, |V_{ts}|)$ can be enhanced as much as $({\mathcal{O}}(\lambda^2), {\mathcal{O}}(\lambda))$ for $|\theta_v|\simeq {\mathcal{O}}(\lambda)$. In such a case, $R$ value can be as low as:  
\be
R=\frac{1}{{\mathcal{O}} (\lambda^2)+1} \simeq  0.95\,.
\ee
Combining Eq. (\ref{eq:fgmax}) with the $4\times 4$ unitarity constraint in Eq.~(\ref{eq:ffunit}), we find that the largest 
possible deviation from the SM value of $V_{ti}$ is obtained for $|\theta_v|\simeq 0.2$ and $|\theta_u|\simeq 0.7$, {\it i.e.},
\be
|V_{td}|\lsim 0.03, \quad |V_{ts}|\lsim 0.2, \quad |V_{tb}|\gsim 0.8\,,
\ee
if we fix the other Wolfenstein parameters in ${\bf V_{3\times 3}^{(0)}}$ as $\lambda=0.22$ and  $A=0.85$. 

\begin{figure}[t]\begin{center}
\psfrag{x}[c][c][0.9]{$\cos\theta_u$}\psfrag{y}[c][c][1]{\rotatebox{270}{$\frac{m_{\tp}}{m_t}$}}
\psfrag{one}[c][c][0.8]{1\ $\sigma$}\psfrag{ninty}[c][c][0.8]{90\% C.L.}\psfrag{nintyfive}[c][c][0.8]{95\% C.L.}
\psfrag{tpmass}[c][c][1]{\rotatebox{270}{$\leftarrow$} }
\psfrag{1.2}[c][c][0.7]{1.2}\psfrag{1.4}[c][c][0.7]{1.4}\psfrag{1.6}[c][c][0.7]{1.6}\psfrag{1.8}[c][c][0.7]{1.8}\psfrag{2}[c][c][0.7]{2.0\ \ }
\psfrag{-0.75}[c][c][0.7]{-0.75}\psfrag{-0.5}[c][c][0.7]{-0.50}\psfrag{-0.25}[c][c][0.7]{-0.25}
\psfrag{0}[c][c][0.7]{0}\psfrag{1}[c][c][0.9]{\ }
\psfrag{0.75}[c][c][0.7]{0.75}\psfrag{0.5}[c][c][0.7]{0.50}\psfrag{0.25}[c][c][0.7]{0.25}
\psfrag{tv1}[c][c][0.8]{{$\theta_v=0.2$}}\psfrag{tv2}[c][c][0.8]{{$\theta_v=0.1$}}
\begin{minipage}{8cm}
\includegraphics[width=8cm]{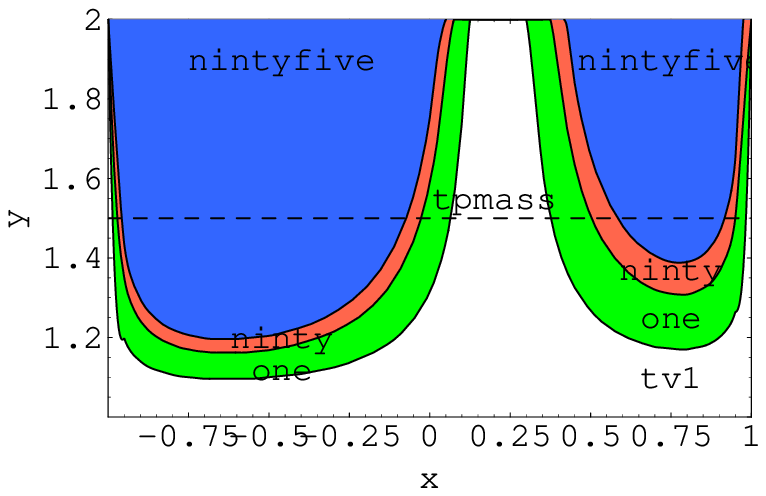}
\begin{center} (a) \end{center}\end{minipage}
\begin{minipage}{8cm}
\includegraphics[width=8cm]{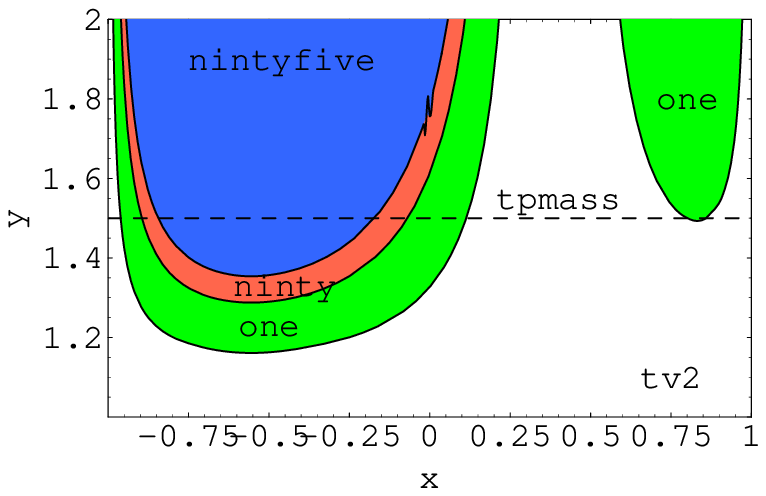}
\begin{center} (b) \end{center}\end{minipage}
\caption{
  Excluded ranges for the mass and mixing of a
  fourth generation $\tp$ quark from the $B\to X_s \gamma $ branching
  ratio at 95\%, 90\% and 68.3\% C.L. The dashed line indicates the experimental bound
  on $m_{\tp}$ (see Eq.~(\ref{eq:tpmass})). We fix the 
  mixing angles  $\theta_w=0$ and $\theta_v\simeq 0.2, 0.1$ for figure (a), (b). Constraints from $R_b$ are similar to those for a vector-like quarks, shown in Fig.~\ref{fig:bsgamma}b.}
\label{fig:bsgfg}\end{center}
\end{figure}

Next, we obtain constraints for $\theta_v$ and $\theta_u$ from a
loop-level process, $B\to X_s \gamma$, by including the $\tp$
contribution. The result is shown in Fig.~\ref{fig:bsgfg}.  In 
Fig.~\ref{fig:bsgfg}a, we fix $\theta_v=0.2$, the maximum allowed value from
the unitarity condition, and find that the allowed range of $V_{tb}$
at $1\sigma$ (95\% C.L.) is $0.07 (-0.07) < V_{tb} < 0.38 (0.58)$.  This
interval does not overlap with the theoretically 
allowed region $|V_{tb}|\gsim 0.71   $, Eq.~(\ref{eq:theory}), and
therefore $|\theta_v|\simeq 0.2$ is excluded.  In Fig.~\ref{fig:bsgfg}b,
where $\theta_v=0.1$, we find that $V_{tb}$ above 0.11 is allowed.

Finally, we consider the constraints from the EW precision
data. The large value of the $S$ parameter in the fourth generation
model is often advocated to exclude this possibility. However, those
analyses are usually performed assuming $T\simeq 0$. As was shown
in the previous section, the $T$ parameter can be modified
significantly in our case due to the mixing between the
fourth generation fermions and the standard fermions (non-zero
$\theta$). Assuming the new $b^{\prime}$ mass equal to the $\tp$ mass,
which ensures a minimal $T$ value, we obtain 
\be T>2.0\sin^2\theta,
\quad U>0.17\sin^2\theta, \quad S>0.16,
\ee 
to be compared with the results of the electroweak fit, $S=0.07\pm 0.10$, and $T=0.13\pm 0.10$ for $U=0$~\cite{EWW}. In fact, a  larger value of $S$ allows a larger value of  
$T$. Thus, this model is still viable for mixing angle and mass configurations similar to the previous model. 

Once again, the ratio $R_b$ turns out to give the strongest constraints. Here, $t$ and $\tp$
loop corrections to $\Gamma (Z\to b\bar{b})$ imply 
\be R_b\lsim
(1-0.019\sin^2\theta) R_b^{\rm SM}.  
\ee This bound, very similar to the one derived for the vector-like $\tp$ case,
Eq.(\ref{eq:rbvt}), requires (at 95\% C.L.)  
\be |\cos\theta_u|\gsim 0.93 
\ee and definitely closes the
unnatural window $|\cos\theta_u|\lsim 1/\sqrt{2}$ left over
by $B\to X_s\gamma$ (see Fig.~\ref{fig:bsgfg}).

We should also mention that gauge anomaly 
cancellation requires the same number of generations in the lepton and quark sectors. 
The fourth generation lepton contributions can also modify the above
predictions quite significantly, depending on their masses (see the detailed
discussion in Ref.~\cite{Holdom:2006mr}).  

\subsubsection{Impact on the single top production}\label{sec:2.2.2}
\begin{figure}[t]
  \hspace*{0.5cm}
  \begin{minipage}[t]{.27\textwidth}
      \epsfig{file=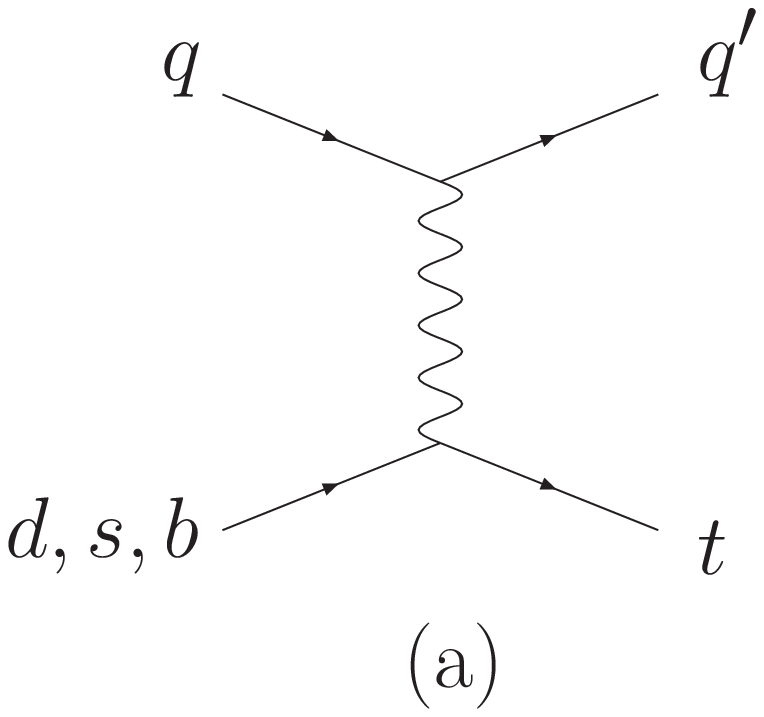, width=3.5cm}
  \end{minipage}
  \begin{minipage}[t]{.27\textwidth}
      \epsfig{file=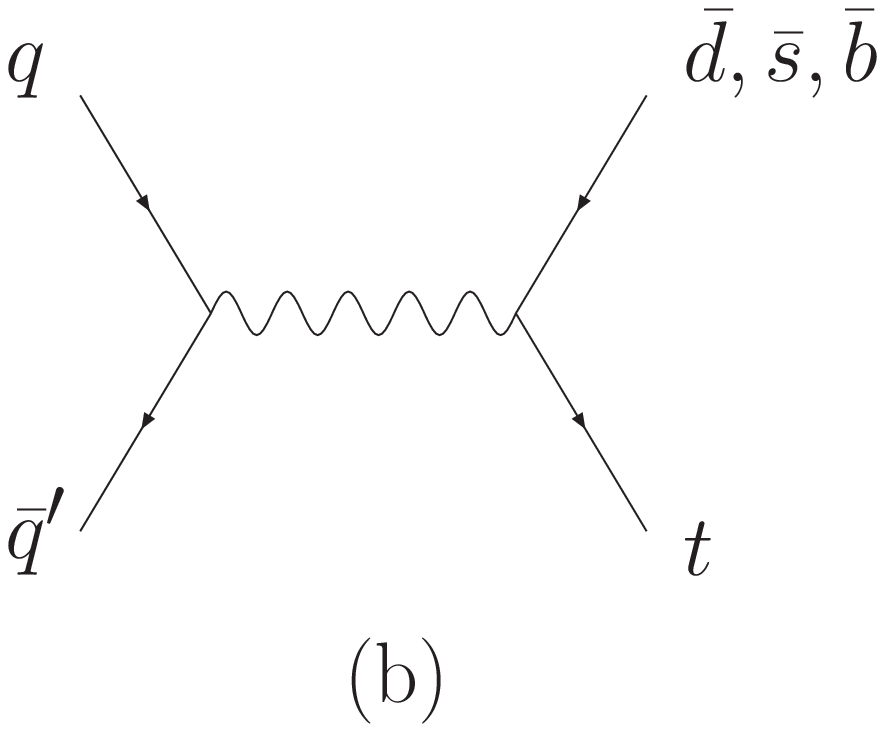, width=3.5cm}
  \end{minipage}
  \hspace*{1cm}
  \begin{minipage}[t]{.27\textwidth}
      \epsfig{file=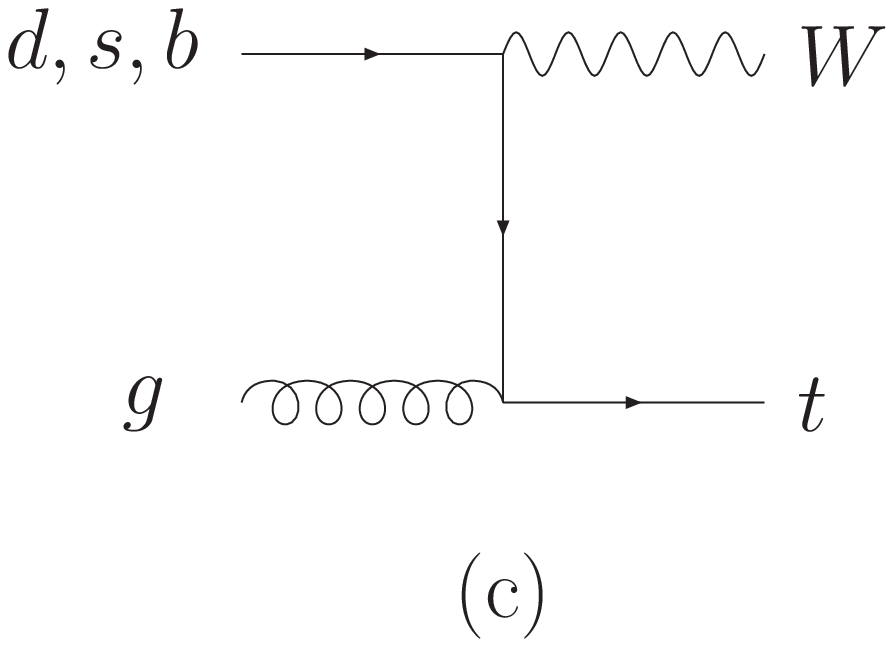, width=3.5cm}
  \end{minipage}
  \caption{Representative diagrams for single top production: $t$-channel (a), $s$-channel (b), and $W$-associated production (c).}
  \label{fig:fynst}
\end{figure}

If  $|V_{td}|$ and  $|V_{ts}|$ are larger than their SM values, a possibility which could 
occur in the fourth generation model but not in the vector-like model, 
both the top branching ratios into $Wj$ and the single top production cross section for
the $t$-channel and $W$-associated production ($Wt$) will be affected (see Fig.~\ref{fig:fynst}). It is interesting to check
what kind of constraints the present limits on the single top production
from the Tevatron give on the $V_{ti}$ matrix elements and what 
the prospects will be at the LHC.  The cross sections for the $t$-channel  production is proportional to the parton distribution functions for the
incoming quark times the corresponding CKM element squared, {\it i.e.}, 
\begin{equation}
\sigma(pp(p\overline{p}) \to tj) = |V_{td}|^2 \sigma_d^{\rm t-ch} + |V_{ts}|^2 \sigma_s^{\rm t-ch} +|V_{tb}|^2 \sigma_b^{\rm t-ch}\,,
\end{equation}
similarly for the $W$-associated production, while the $s$-channel can be written as:
\begin{equation}
\sigma(pp(p\overline{p}) \to tq; \ \ q=d,s,b ) = (|V_{td}|^2 + |V_{ts}|^2 +|V_{tb}|^2 )\, \sigma^{\rm s-ch}\,.
\end{equation}
In Table~\ref{tab:topx} the results for the cross sections calculated
at LO with {\tt MadGraph/MadEvent}~\cite{MGME} ($m_t=175$ GeV,
$\mu_R=\mu_F=m_t$, PDF=CTEQ6L1~\cite{Pumplin:2002vw}) at the Tevatron
and LHC are given as coefficients of the corresponding CKM matrix
element. If the three-family unitarity holds, the contributions
coming from the strange and down quarks
are suppressed by the smallness of the corresponding CKM elements and give a 
negligible contribution to the total cross section. 
\begin{table}[t]
\begin{center}
\begin{tabular}{|c|c|ccc|}
\hline
Collider   & Process   &\multicolumn{3}{|c|}{Cross section (pb)} \\ \hline
\multicolumn{2}{|c|}{ }&$|V_{tb}|^2$ & $|V_{ts}|^2$ & $|V_{td}|^2$ \\ \hline
   &$t$-channel&    0.88         &     2.7         &          10.5          \\
 Tevatron           &$s$-channel&       \multicolumn{3}{|c|}{0.30 }\\
           &$Wt$       &   0.038     &  0.150            & 1.26                    \\
\hline
                    &$t$-channel&     150(87)     & 277   (172)        &  766  (253)              \\ 
LHC         &$s$-channel&      \multicolumn{3}{|c|}{ 4.6 (3.4) }                    \\
            &$Wt$       &      30      &   67        &    294 (107)             \\     
\hline
\end{tabular}
\end{center}
\caption{Contributions to the cross section for single top 
production proportional to the corresponding CKM element squared. 
Cross sections (in pb) are calculated at LO 
($m_t=175$ GeV, $\mu_R=\mu_F=m_t$, PDF=CTEQ6L1~\cite{Pumplin:2002vw}) and refer to the 
production of a top. The anti-top cross sections are given 
in parenthesis when different from those of a top.}
\label{tab:topx}
\end{table}

The above predictions can be compared to the most stringent limits from 
the CDF collaboration~\cite{CDF8185}: \\
\bea
\sigma^{\rm s-ch}_{\rm SM}+\sigma^{\rm t-ch}_{\rm SM} &<& 3.4\ {\rm pb\ at}\ 95\%\ {\rm C.L.} \nonumber \\
\sigma^{\rm s-ch}_{\rm SM} &<& 3.1\ {\rm pb\  at}\ 95\%\ {\rm C.L.} \label{eq:CDFsingle}\\
\sigma^{\rm t-ch}_{\rm SM} &<& 3.2\ {\rm pb\  at}\ 95\%\ {\rm C.L.}  \nonumber
\eea
These limits assume a SM scenario, with $V_{tb}=1$.  In order to
curb the large background coming mainly from $W+jets$ and $t\bar t$,
the experimental analysis makes extensive use of the kinematical
information of the signal, such as the presence of forward jet and/or
of a charge asymmetry in the $t$-channel. However, the most important
selection criterium is given by the requirement of two jets, of which
one or two are $b$-tagged. If $V_{tb}=1$, the $t$-channel typically
leads to one $b$-jet in the final state (from the top decay), while
the $s$-channel to two $b$-jets. For sake of argument we restrict the
following study to this distinctive feature, keeping in mind that the
results obtained here are meant as illustration and could be
easily improved by a more detailed analysis.

\begin{figure}[t]
\psfrag{Vtd}[c][c][0.8]{$|V_{td}|$}
\psfrag{Vts}[c][c][0.8]{$|V_{ts}|$}
\psfrag{Vtb}[c][c][0.8]{$|V_{tb}|$}
\psfrag{0.8}[c][c][0.7]{0.8}
\psfrag{0.6}[c][c][0.7]{0.6}
\psfrag{0.4}[c][c][0.7]{0.4}
\psfrag{0.2}[c][c][0.7]{0.2}
\psfrag{1}[r][c][0.7]{1.0}
\psfrag{R}[c][c][.9]{$R$}
\psfrag{ST}[c][c][.9]{\small single top}
\psfrag{physical}[c][c][.9]{\small physical $\longrightarrow$}
\psfrag{all}[c][c][.9]{\small combined $\to$}
\begin{center}
\vspace*{.2cm}
\hspace*{0cm}
\begin{minipage}{16cm}
\epsfxsize=5.1cm \epsfbox{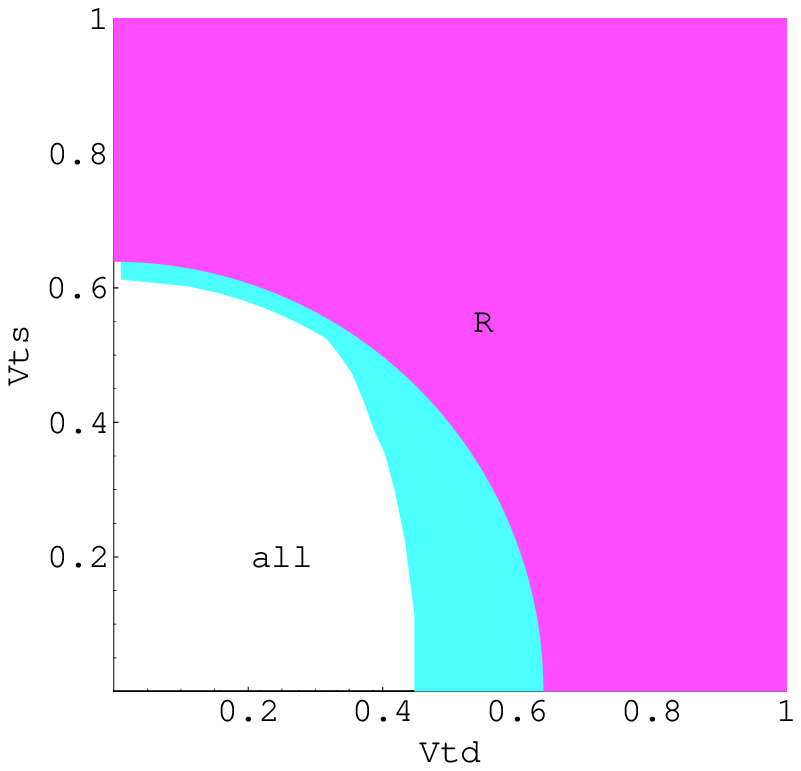}\ 
\epsfxsize=5.1cm \epsfbox{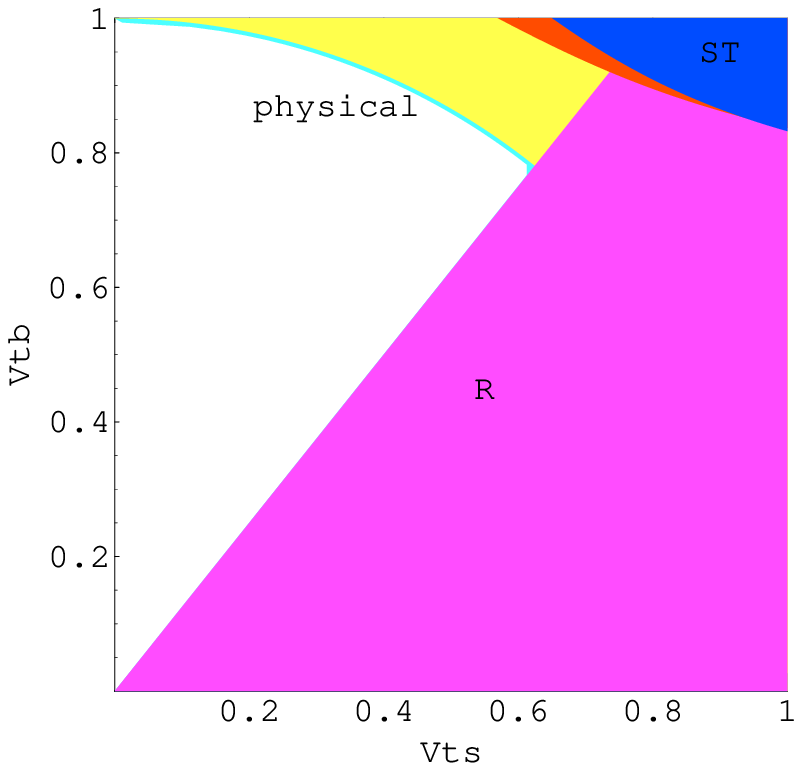}\ 
\epsfxsize=5.1cm \epsfbox{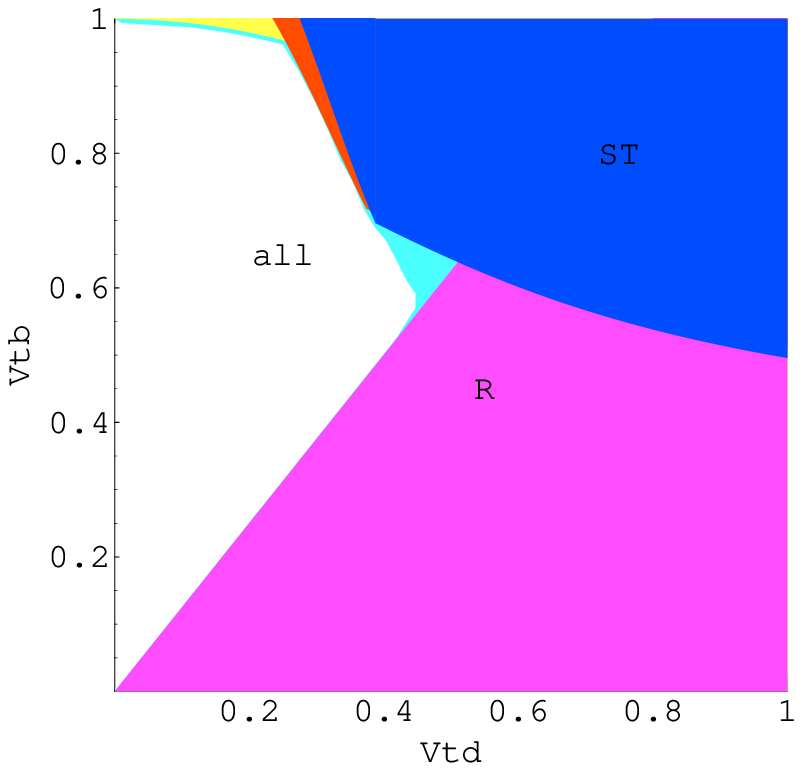}\ 
\end{minipage}
\end{center}
\caption{Excluded regions for $|V_{td}|$, $|V_{ts}|$, and $|V_{tb}|$ as obtained from the CDF limits 
on the single top production, $\sigma_{1b}$ (orange) and
$\sigma_{1b}+\sigma_{2b}$(dark-blue), the measurements of $R$ (pink)
and the physical bound $|V_{td}|^2+|V_{ts}|^2+|V_{tb}|^2<1$ (yellow).
The combination of these four bounds provides an additional excluded
region (light-blue).  } \label{fig:CDFcon}
\end{figure}

In this approximation, the limits on $\sigma^{\rm s-ch}_{\rm SM}$ and  $\sigma^{\rm t-ch}_{\rm SM}$
can be translated into the cross section involving one $b$-jet, $\sigma_{1b}$,  
and two $b$-jets $\sigma_{2b}$ and their sum, 
$\sigma_{tot}=\sigma_{1b}+\sigma_{2b}$, where 
\begin{eqnarray}
\sigma_{1b}&=& R \left\{ 2 (|V_{td}|^2+|V_{ts}|^2) \sigma^{\rm s-ch}+ 
[|V_{td}|^2\sigma_{d}^{\rm t-ch}+
 |V_{ts}|^2\sigma_{s}^{\rm t-ch}+
 |V_{tb}|^2\sigma_{b}^{\rm t-ch}] \right\}\,,\\
\sigma_{2b}&=& R \,|V_{tb}|^2\, \sigma^{\rm s-ch}. 
\end{eqnarray}
$R$ is defined in Eq.~(\ref{eq:Rdef}).
Using the constraints in Eq. (\ref{eq:CDFsingle}) and the result $R>0.61$ at 95\% C.L., we obtain the excluded regions for $|V_{ti}|$ as 
shown in Fig.~\ref{fig:CDFcon}. The resulting allowed values,  $|V_{td}|\lsim 0.46$ and $|V_{ts}|\lsim 0.62$, are much less constrained than those obtained from the $4\times 4$ unitarity and $B\to X_s\gamma$. 

\section{Future prospects at the LHC}\label{sec:3}

In this section we discuss the perspectives for the determination of
$V_{tb}$ at the LHC.  The primary method to extract information on
$V_{tb}$ will be through the measurement of the single top cross
sections, which are directly proportional to $|V_{tb}|^2$. 
The best determination will come from $t$-channel production, 
but it will still be crucial to have measurements from all the three channels
to identify possible sources of new physics, since
in general new models may have effects in one
channel and not in the others~\cite{Tait:2000sh}. For the models
introduced in the previous section, it will also be possible to study
the production of extra heavy quarks and from that to discriminate,
for instance, the case of just one vector-like top from that of a full
$SU(2)_L$ doublet. We briefly illustrate this possibility and outline
possible strategies in Section~\ref{sec:tp}. We mention in passing that
another handle to $V_{tb}$ might be offered by the direct measurement of
the top width. There have been suggestions on how to perform such a
measurement in $e^+e^-$ experiments~\cite{Orr:1993kd,Batra:2006iq}. 
We do not discuss this possibility here, even though such studies at 
the hadron colliders would be certainly welcome.

\subsection{$V_{tb}$  measurement at the LHC}
Going from Tevatron to LHC, the higher energy and luminosity provide
better possibilities for a precise determination of the CKM matrix
element $V_{tb}$, in all the three production modes:
$t$-channel ($q^2_{\mathrm{W}}<0$), $s$-channel
($q^2_{\mathrm{W}}>0$), and  $W$-associated production
($q^2_{\mathrm{W}}$=$M^2_{\mathrm{W}}$). The corresponding cross
sections are shown in Table~\ref{tab:xsections}~\cite{Smith:1996ij,Stelzer:1997ns,Campbell:2005bb}.
\begin{table}[h]
\begin{center}
\begin{tabular}{|c | c|}
\hline
Process & $\sigma$ (pb) \\
\hline
$t$-channel & 245 \\
$Wt$ & 60 \\
$s$-channel & 10\\
\hline
\end{tabular}
\end{center}
\caption{\label{tab:xsections} The single top production cross section values at 
the LHC at the NLO level (top and anti-top contributions are summed).}
\end{table}
The three production processes occupy different phase space regions
and have large differences in signal-to-background ratios.

\subsubsection{Determination  of $V_{tb}$ from the $t$-channel production}

For the $t$-channel, the signature is one lepton, missing energy, one
$b$-jet and one recoil jet (un-tagged and at high rapidity). In the
CMS study of Ref.~\cite{ref:ts} it is shown that a
signal-to-background ratio higher than unity is achievable and the
main background after selection is $t\bar t$.

\begin{figure}[t]
\begin{center}
\rotatebox{0}{\includegraphics[width=11cm]{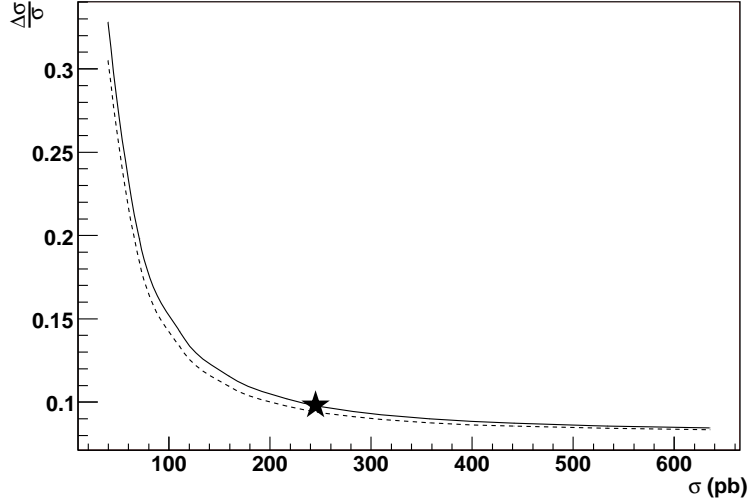}}
\end{center}
\caption{ The relative uncertainty  on the cross section as a function of the cross section
for the $t$-channel, corresponding to 10 fb$^{-1}$ integrated luminosity (solid line).
The star indicates the SM expectation. 
The dashed line represents the systematic uncertainty.   
}
\label{fig:tchannel}
\end{figure}

The total relative uncertainty  on the cross section can be estimated by:
\begin{eqnarray}
\frac{\Delta\sigma}{\sigma } = \frac{\sqrt{N_S+N_B}}{N_S} 
\oplus \frac{\Delta N_S\oplus \Delta N_B}{N_S} \oplus \frac{\Delta L}{L}, \label{eq:err}
\end{eqnarray}
where $N_S$ and $N_B$ are the number of selected signal and background
events respectively, and $L$ and $\Delta L$ are the LHC luminosity and
its uncertainty.  $\Delta N_S$ and $\Delta N_B$ are the experimental systematics (such as uncertainties on jet energy scale and $b$-tagging efficiency)
for the signal and the background, respectively. In the latter 
the uncertainty on the background sample normalization is also included.  
Fig.~\ref{fig:tchannel} shows its
dependence on the signal cross section.
For 10 fb$^{-1}$ of integrated luminosity and under the assumption that
the signal cross section is as expected in SM, this results
in\footnote{In Ref.~\cite{ref:ts},  8\% systematics  is quoted
because it includes 4\% uncertainty on $\sigma^{\rm th}$ which we
add separately later in this section.}
\begin{eqnarray}
\frac{\Delta\sigma}{\sigma} & = & \pm 3\%{\rm (stat.)}\pm 7\%{\rm (syst.)}\pm 5\%{\rm (lum.)}.
\end{eqnarray}
 The measurement is systematics dominated, mostly due to 
the imperfect knowledge of jet energy scale,   $b$-tagging efficiency and mistag probability. 

The expected uncertainty  on $V_{tb}$ may be computed as
\begin{eqnarray}
\frac{\Delta V_{tb}}{V_{tb}} & = & 
\frac{1}{2}\left(\frac{\Delta \sigma^{\rm meas}}{\sigma^{\rm meas}} 
\oplus \frac{\Delta \sigma^{\rm th}}{\sigma^{\rm th}}\right).
\end{eqnarray}
For the $t$-channel, the uncertainty  on $\sigma^{\rm th}$ has been 
calculated in detail in Ref.~\cite{Sullivan:2004ie} and has 
the following contributions: 
\begin{itemize}
\item PDF uncertainties: $+1.3\%, -2.2\%$,
\item higher orders (QCD scale): $3\%$,
\item variation of the top mass within 2 GeV:  $+1.56\%, -1.46\%$,
\item uncertainty on the b-quark mass: $< 1 \%$. 
\end{itemize}

The above uncertainties are associated to the fully inclusive cross
section.  Therefore, the overall uncertainty on $V_{tb}$ is estimated
to be 5\%. A more accurate determination would take into account the
specific phase space region selected by the analyses.  In particular,
we point out that the request of exactly two jets (vetoing any other
jet above a certain threshold), needed to reduce the $t\bar t$
background to a reasonable level, may give a larger scale dependence
than quoted above.

Moreover, more studies are needed on the electroweak corrections. Due
to the presence of the $W$ in the intermediate state, real and virtual
photon emissions are expected to give sizable amplitudes, and the
correction to $\sigma^{\rm th}$ might be 
as large as several percents~\cite{Beccaria:2006ir}.

\subsubsection{Other single top processes}

For the $W$-associated production, one can follow two
complementary search strategies: one based on the selection of two isolated
leptons, the other with one isolated lepton and two light jets
compatible with the $W$ mass. In both cases missing energy and one
$b$-jet are also required in the final state, and no other jet is
allowed. The main limitation of this analysis is the similarity of the
signal with the $t\bar t$ background, where the jet counting is the
only handle to reduce it. It is worth mentioning that such a similarity
with the $t\bar t$ is also a problem at the theoretical level: $Wt$ is
consistently defined and insensitive to the quantum interference with
$t\bar t$ only when extra $b$-jets in the final state are
vetoed~\cite{Campbell:2005bb}.

After the selection, a signal-to-background ratio of 0.37 is expected
for the di-leptonic channel and 0.18 for the single-leptonic, the
background being almost completely constituted by $t\bar t$ events.
In order to constrain this background, and to cancel out a large part
of the main systematics, one can make use of a control sample and employ
the so-called ``ratio method''~\cite{ref:wt}. Then, the cross 
section can be rewritten  as 
\be \sigma^{Wt} =
\frac{R_{tt}(N-B_0)-(N_c-B_c)}{\epsilon^{Wt}(R_{tt}-R_{Wt})} \ee where
$N (B_0)$ and $N_c (B_c)$ are the total number of selected events (the
non-$t\bar{t}$ background) in the main and in the control samples,
respectively.  $\epsilon^{Wt}$ is the signal selection efficiency.
$R_{Wt} (R_{tt})$ is the ratio of the efficiency in the control sample
to the efficiency in the main sample for the signal (and $t \bar t$). The 
uncertainty in the background sample normalization,
which dominates $\Delta N_B$ in Eq. (\ref{eq:err}), is now
associated to the statistical uncertainty  in the large control sample of $N_c$
and the systematic uncertainty  due to the background rejection is highly
reduced since it only enters in the ratio.

The expected precision on the cross section  with 10 fb$^{-1}$ of integrated luminosity is: 
\be
\frac{\Delta\sigma}{\sigma}  = \pm6\%{\rm (stat.)}\pm16\%{\rm ( syst.)}\pm5\%{\rm (lum.)}.  
\ee
This result is obtained by averaging di-leptonic and single-leptonic analyses from  Ref.~\cite{ref:wt} assuming fully correlated systematic uncertainty. 
The statistical significance for 10 fb$^{-1}$ is higher than six standard deviations.

Although not competitive with the $t$-channel production in terms of
the achievable precision in the extraction of $V_{tb}$, the
$W$-associated process is still attractive since the observation of
the $W$ in the final state would prove that the top is produced
through a charged current interaction. As we mentioned above, the
definition and the measurement of this channel is difficult due to the
large overlap in phase space with $t\bar t$, whose cross section is
more than ten times larger.  In this respect it is interesting to note
that in $\gamma p$ collisions at the LHC, where protons emit almost real photons colliding with protons of the opposite beam,  the $Wt$
and $t\bar t$ cross sections are of a similar size, leading to a much
better signal over background ratio. Work to explore this alternative
is on-going~\cite{CP3-gammap}.

For the $s$-channel process $q\bar q'\to W^{*} \to t\bar b/\bar tb$, whose signature is
one lepton, missing energy and two $b$-jets, the $t\bar t$ background
is again difficult to curb and a ratio method has to be applied as in
the $Wt$ case.  
The final result of the analysis~\cite{ref:ts}, for 10 fb$^{-1}$, is:
\begin{eqnarray}
\frac{\Delta\sigma}{\sigma} & = & \pm 18\%{\rm (stat.)}\pm 31\%{\rm (syst.)}\pm 5\%{\rm (lum.)},
\end{eqnarray}
where most of the contribution to the systematics  comes from the jet energy scale uncertainty.

\subsection{$\tp$ production cross sections at the LHC}
\label{sec:tp}

If extra quarks exist, either as a $SU(2)$ gauge singlet or in a
doublet, and they are light enough, they could be also discovered at
LHC.  The phenomenology of such states has been studied in the
literature (see, {\it e.g.}, Refs.~\cite{Aguilar-Saavedra:2005pv,Han:2003wu}) 
and here we limit ourselves to a brief discussion, highlighting 
how the $SU(2)$ nature of the extra quark(s) could be determined.

\begin{figure}[t]
  \hfill
  \begin{minipage}[t]{.48\textwidth}
    \begin{center}  
      \epsfig{file=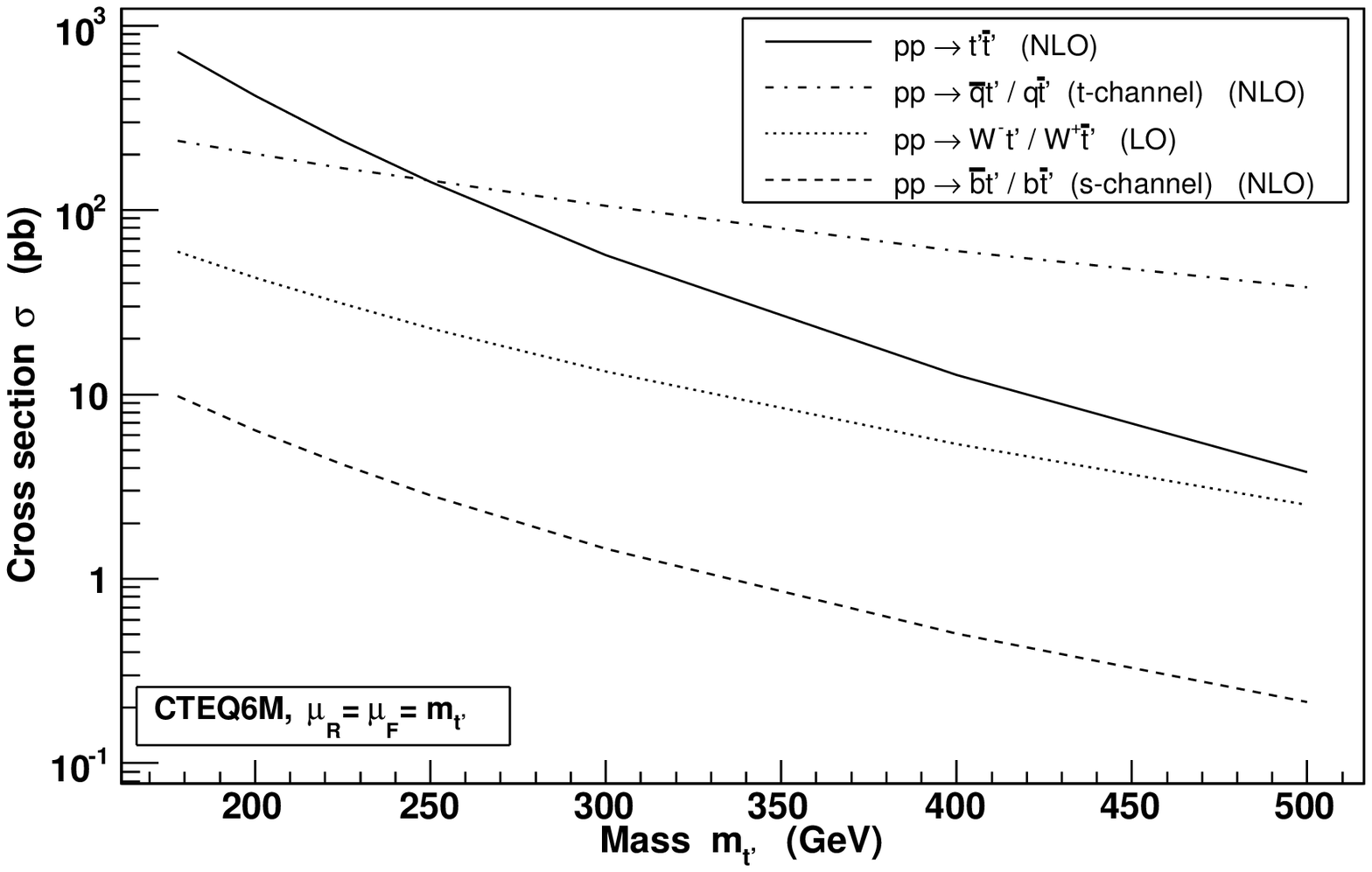, width=1.1\textwidth}
      \caption{$t'$ production as a function of its mass, with $|V_{bt'}|$ and $|({\bf VV^{\dagger}})_{tt'}|$ set to one. 
Results are shown for $t'\bar t'$ pair production and the three single $\tp$ channels.}
      \label{extra_top_1234}
    \end{center}
  \end{minipage}
  \hfill
  \begin{minipage}[t]{.48\textwidth}
    \begin{center}  
      \epsfig{file=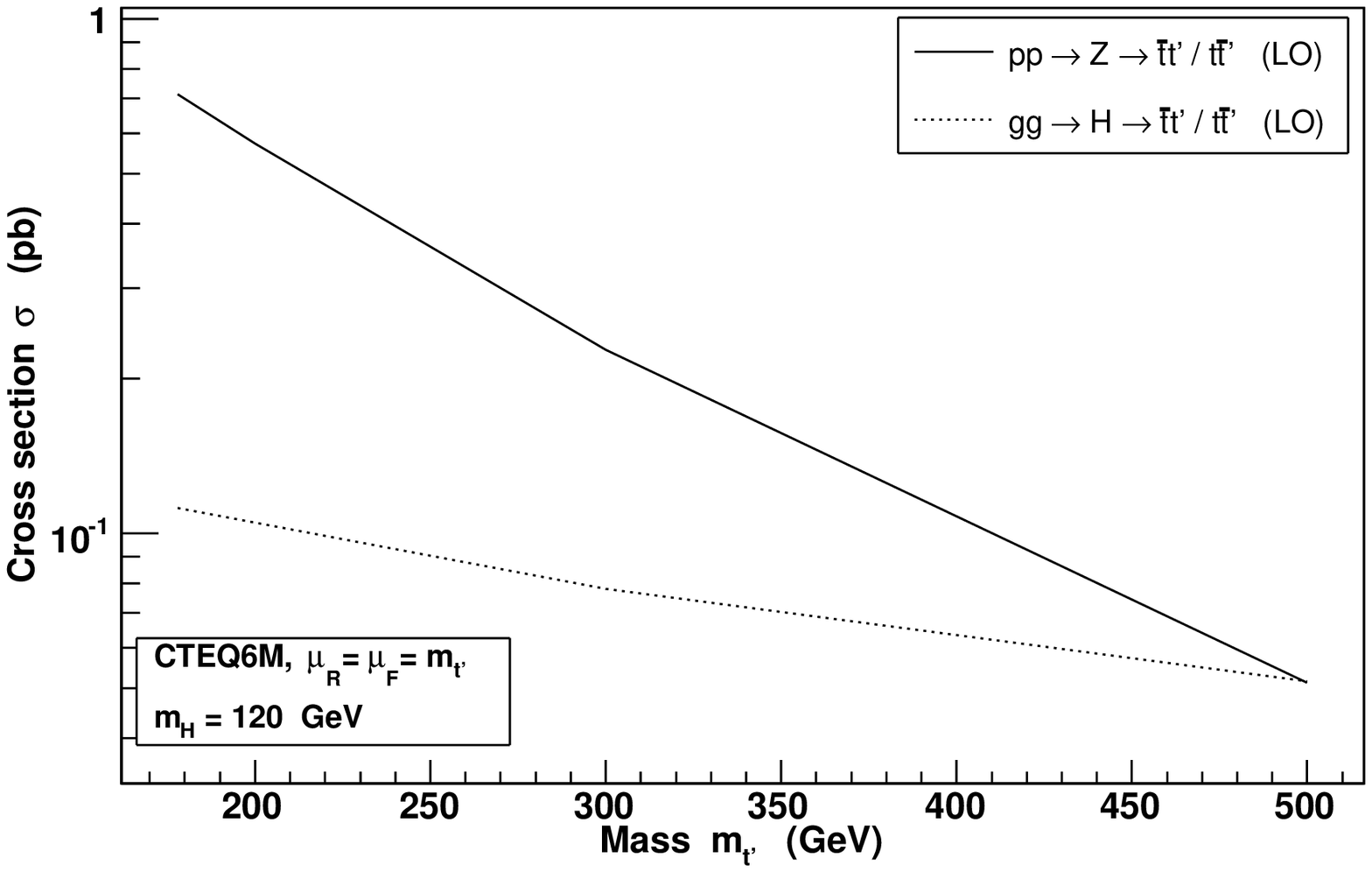, width=1.1\textwidth}
      \caption{FCNC $\bar tt'/t \bar t'$ production through an $s$-channel $Z$ or Higgs boson (solid and dotted lines) as a function of the mass of the $t'$, with $|({\bf VV^{\dagger}})_{tt'}|$ set to one.}
      \label{extra_top_678}
    \end{center}
  \end{minipage}
  \hfill
\end{figure}

In Figs.~\ref{extra_top_1234} and~\ref{extra_top_678} 
the $t'$ production cross sections are shown for various production modes as a
function of the $t'$ mass. For simplicity, we have set $|V_{\tp b}|=
|({\bf VV^{\dagger}})_{tt'}|=1$, so that if the mass of the $t'$ is
equal to the top mass ($\sim 175$ GeV) the cross sections are equal to
the SM cross sections for top production. Results at LO
have been obtained with {\tt MadGraph/MadEvent}~\cite{MGME}, while {\tt MCFM}~\cite{Campbell:1999ah} has been used when calculations at next-to-leading order in QCD were 
available. 
 
In Fig.~\ref{extra_top_1234} the double $t'$ production cross
section is given by the solid line and the  single $t'$ production channels 
are given by the dashed ($s$-channel), dash-dotted ($t$-channel) and dotted ($W\tp$)
lines. For $t'$ masses below $\sim 250$ GeV double $t'$
production dominates the single $t'$ production, just as the double
top cross section is larger than the single top in SM.
Above $\sim 250$ GeV the $t$-channel becomes the dominant 
production mechanism, as it is the least dependent on the $t'$
mass. Note, however, that the single $t'$ production scales as $|V_{\tp
b}|^2$, while the pair production cross section is independent of it and
might still be the dominant production mechanism. For example, for
$\cos\theta=0.71$ the single $t'$ production cross sections decrease
by an overall factor of four.

One way to distinguish between a new extra doublet and a vector-like
quark is to look for FCNCs, which are only present for the vector-like case. 
At leading order there are two mechanisms
for the production of a $t\bar{t}'/t' \bar{t}$ pair, {\it viz.}, 
through an $s$-channel $Z$ or Higgs boson. 
The total cross section for the processes
$pp\rightarrow Z \rightarrow  t\bar{t}'/ \bar{t}t' $ and $gg\rightarrow
H \rightarrow \bar{t}t' / t\bar{t}'$ are given by the solid and the
dotted lines in Fig.~\ref{extra_top_678}, respectively. 
Note that the $gg\rightarrow H \rightarrow\bar{t}t' / t\bar{t}'$ cross section is almost independent of the $t'$
mass because of the cancellation of two competing effects, {\it i.e.}, 
the increase of the $tt'H$ coupling and the gluon luminosity
suppression for larger $x$.

\section{Conclusions}

In this paper we have elaborated on the phenomenology concerning the 
CKM matrix element $V_{tb}$ in models that relax the strong constraints
coming from the unitarity. We have first emphasized that $V_{tb}\simeq 1$ is 
required neither from $B$ physics nor from the top quark decay rate 
measurements. Only the direct extraction of $V_{tb}$ from the single top
production cross section at the Tevatron and at the LHC 
will allow to complete our knowledge of the CKM matrix and hopefully 
shed new light on the nature of the top quark.

As a simple extension of the SM that breaks the $3\times 3$ unitarity
condition of the CKM matrix and leads to a deviation from $V_{tb}\simeq
1$, we have considered the addition of extra fermions: either a vector-like
up-type quark ($\tp$) or fourth generation quarks ($\tp$ and
$b^{\prime}$). The main motivation for selecting these models is that
they serve well the illustrative purpose of our study. 
They are simple, self-consistent and allow to easily  find
the constraints on $V_{tb}$ coming both from precision physics and 
direct observation. In this respect, they should be regarded as useful
templates for further experimental scrutiny on $V_{tb}$. 

We find that the strongest constraint on these models comes from
$R_b$, which severely restricts the allowed amount of $t-t'$ mixing. When this
result is combined with the very recent direct bound on the $\tp$ mass
by the CDF collaboration, $m_{\tp}\gsim 258$ GeV, one finds
$|V_{tb}|> 0.9$. This very strong bound relies, however, on two
assumptions which  might not hold in more sophisticated models.  The first
one is that the corrections to $R_b$ induced by loop effects are only
coming from the $\tp$ contribution, and therefore models with an
extended particle content may be less  constrained. The
second assumption,  which is at the basis of the lower bound on the $t'$
mass by CDF, is that the branching ratio of $t' \to W q $ is one. For
instance, this condition is satisfied in our vector-like $t'$ model
only for $m_{t'}\lsim 300$ GeV. If at least one of the above conditions
is not fulfilled, we have shown that other indirect measurements, such as 
those coming from 
 $B\to X_s\gamma$ or  
the $S,T,U$ oblique parameters should also be considered.

In the near future the observation of the single top process,
which is challenging both at the Tevatron and the LHC, will for the first time provide 
 a direct measurement of $V_{tb}$.  We showed that the
current lower bound from Tevatron data has started giving direct
information on the magnitudes  of $V_{td}$ and $V_{ts}$, and that they will
be further constrained as soon as the LHC data will be
available. Among all three possible production mechanisms, the
$t$-channel is the most promising process where $V_{tb}$ could be
determined at the 5\% precision level already with  10 fb$^{-1}$
of integrated luminosity. The
precision of this result is limited by the systematic uncertainty and might be well improved with better understanding of the detector and background.  The other channels,
$W-$associated and $s$-channel, are more challenging due to a much
larger systematic uncertainty. However, a measurement of these production
mechanisms will be important to  complete our knowledge  of the top
quark coupling to the weak current and possibly reveal new physics.

\bigskip
\section*{Acknowledgments}
We are thankful to Tony Liss and Scott Willenbrock for discussions. 
We thank the organizers of the CERN workshop ``Flavour in the era of the LHC''  for
the nice atmosphere that stimulated this work. The work was supported by 
the Belgian Federal Office for Scientific, Technical and Cultural Affairs 
through the Interuniversity Attraction Pole P5/27. 


\end{document}